\documentclass[12pt]{article}

\usepackage{times}
\input epsf
\epsfclipon
\usepackage{epsfig}
\usepackage[numbers]{natbib}
\usepackage{amssymb}
\usepackage{amsthm}
\usepackage{mathtools}

\usepackage{soul}

\title{From equality to diversity - bottom-up approach for hierarchy growth}

\author{Agnieszka Czaplicka$^{1,2}$, Janusz A. Ho{\l}yst$^{1}$ \\
\\
\normalsize{$^{1}$Faculty of Physics, Warsaw University of Technology,}\\
\normalsize{Koszykowa 75, Warsaw, 00-662, Poland.}\\
\normalsize{$^{2}$Center for Humans and Machines, Max-Planck Institute for Human Development,}\\
\normalsize{Lentzeallee 94, Berlin 14195, Germany.}
}

\begin{document}

\maketitle

\begin{abstract}
Hierarchical topology stands as a fundamental property of many complex systems. In this work, we present a simple yet insightful model that captures hierarchy growth from bottom to top. Our model incorporates two key dynamic processes: the emergence of local leaders through promotions, where successful agents advance to higher hierarchical levels by attracting followers, and the natural degradation of agents to the lowest level. From an initial flat structure where all agents occupy the bottom level, the system evolves toward a stationary state characterized by an exponential distribution of agents across levels—a pattern remarkably similar to those observed in diverse real-world hierarchies, from hunter-gatherer societies and mammalian groups to online communities. Notably, while the average hierarchy level and the fraction of ground-level agents remain independent of system size, the maximum height of the hierarchy grows logarithmically with the total number of agents. In the stationary state, agents maintain a significantly smaller number of followers compared to their peak influence at the promotion moment. Results from numerical simulations are supported by analytical solutions derived based on the rate equations.
\end{abstract}



\section{Introduction}
\label{sec:Introduction}


Hierarchical structures are pervasive in our natural, social and technological environments \citep{pumain,zafeiris2017we,dominance, met1,prot1,www,soc1,soc4,huamnhier}. The concept of hierarchy \citep{pumain4} encompasses several distinct properties of complex structures: $(i)$ the ordering of system elements according to increasing or decreasing values of a selected  variable \citep{pumain4,pumain6}, such as city size  or human age (order hierarchy); $(ii)$ inclusion relations in nested topologies \citep{simon1,simon2}, as seen in academic structures where faculties contain institutes, which in turn contain laboratories (inclusion hierarchy); $(iii)$ the emergence of higher-level structures  through  interactions between lower-level elements \citep{simon1,simon2,anderson2,holland}, e.g. cells forming organs and organs forming animals (level hierarchy); and $(iv)$ the control of lower-level members by higher-level elements, as observed in management structures (control hierarchy).     

These different kinds of hierarchical systems arise for different reasons \citep{pumain}. For example, the control hierarchy, especially when expressed as social heterogeneity, facilitates efficient division of tasks / roles among community members and encourages specialization \citep{vicsek}. Crucially, the structure of biological and social groups is neither flat nor static, evolving over time even in the absence of external stimuli. As a consequence, when modelling the emergence of hierarchies, we must consider different scenarios of system dynamics, with {\it initial conditions} playing a pivotal role. Two contrasting scenarios are particularly noteworthy: {\it bottom-up} formation \citep{bonabeau,cedric,ieee}, where the initial state is a homogeneous phase, and {\it top-down} evolution, where development begins with a charismatic leader \citep{kac1,kac2,kac3} who serves as the highest (and permanent) authority for the growing community \citep{pre}.
    
Political and social organisations are examples of hierarchical communities that usually emerge from the bottom-up from initially equal peer agents. These highly structured societies are built through complex interactions between individuals, each pursuing their own goals and strategies for success. This pursuit generates various competitive and cooperative behaviours, leading to the formation of diverse groups composed of agents with shared views by their political preferences, economic interests, or religious beliefs. As groups grow, direct interactions between all members become inefficient, necessitating the selection of representatives (spokespersons, leaders) to manage the intergroup interactions. The hierarchical organization of society naturally results from this process. There are numerous illustrations in the real world of this form of social organisation. It can be observed among Wikipedia editors \citep{Lerner_2017}, developers of open source software \citep{opensource}, and members of various cultural associations \citep{cop}. Other examples include hunter-gatherer groups \citep{huntergatherer}, mammalian societies \citep{mammals}, and human groups in virtual world \citep{fuchs}.

Biologists have extensively studied various hierarchies in animal groups. Bonabeau et al. \citep{bonabeau} developed a model of self-organizing hierarchies in animal societies, where evolution is driven by continuous pairwise competitions, with an individual's hierarchical position and strength determined by their success in these encounters. This model has been  thoroughly investigated through numerical simulations and non-equilibrium statistical physics methods \citep{stauffbon,malarzbon}. An alternative framework was proposed by Nepusz and Vicsek \citep{vicsek}, in which hierarchical structures emerge through bottom-up processes, as less successful individuals imitate the behaviour of more accomplished group members.
In our previous work \citep{pre}, we examined a model of evolving control hierarchy initiated from a top node (leader) and governed by tournament selection rules. Our findings revealed that with a fixed tournament size, the number of hierarchical levels grows continuously. However, when tournament size is proportional to system size, the number of levels converges to a limiting value, an effect attributed to the available information during system growth.

In this study, we investigate a model complementary to our previous work by examining bottom-up system growth, where hierarchical structures develop upward from a foundation of independent individuals. We explore how self-organization processes transform a set of equal peers into complex hierarchical structures, focusing on collaborative tendencies and social hierarchy formation. Our model represents inter-agent dependencies through links between adjacent levels and differs from previous studies by emphasizing the collective nature of hierarchical development, where one agent's promotion can cascade through their network of influence. Additionally, we incorporate agent degradation to model the natural turnover of experienced members with newcomers. Our investigation combines qualitative and quantitative analyses of system evolution, supported by both computer simulations and analytical description. Hierarchical systems described here possess features of control hierarchy (by leader-follower relation) and ordering hierarchy (hierarchy level corresponds to higher position in society).

\section{Algorithm of hierarchy growth}
\label{sec:Algorithm}
In this study, we consider a group of $N$ individuals (nodes or agents) that are initially independent (at time $t=0$). In the course of time peer agents establish connections leading to the emergence of hierarchical levels. The links between agents correspond to interactions that result in the formation of hierarchies akin to those observed in the real world. It is important to note that although we display only 'vertical' connections between agents we also assume 'horizontal' interactions among nodes at the same level which are revealed through promotion processes.  The hierarchy level $h$ occupied by an agent is interpreted as their position (relevance) in society, where the relevance of an agent is assumed to increase with the value $h$. Hierarchy growth proceeds from the bottom (ground level $h=0$) to the top. Let us consider two connected agents $i$ and $j$ at levels  $h_i=h_j+1$. This kind of dependency will be treated as a \textit{leader} (agent $i$) - \textit{follower} (agent $j$) relation. Please note that there are no direct connections between followers of one leader. Furthermore, it was assumed that each agent can have no more than one leader. Hence, our network is free of cycles and can be interpreted as a tree-like topology. 

\begin{figure}[ht]
\centerline{\psfig{file=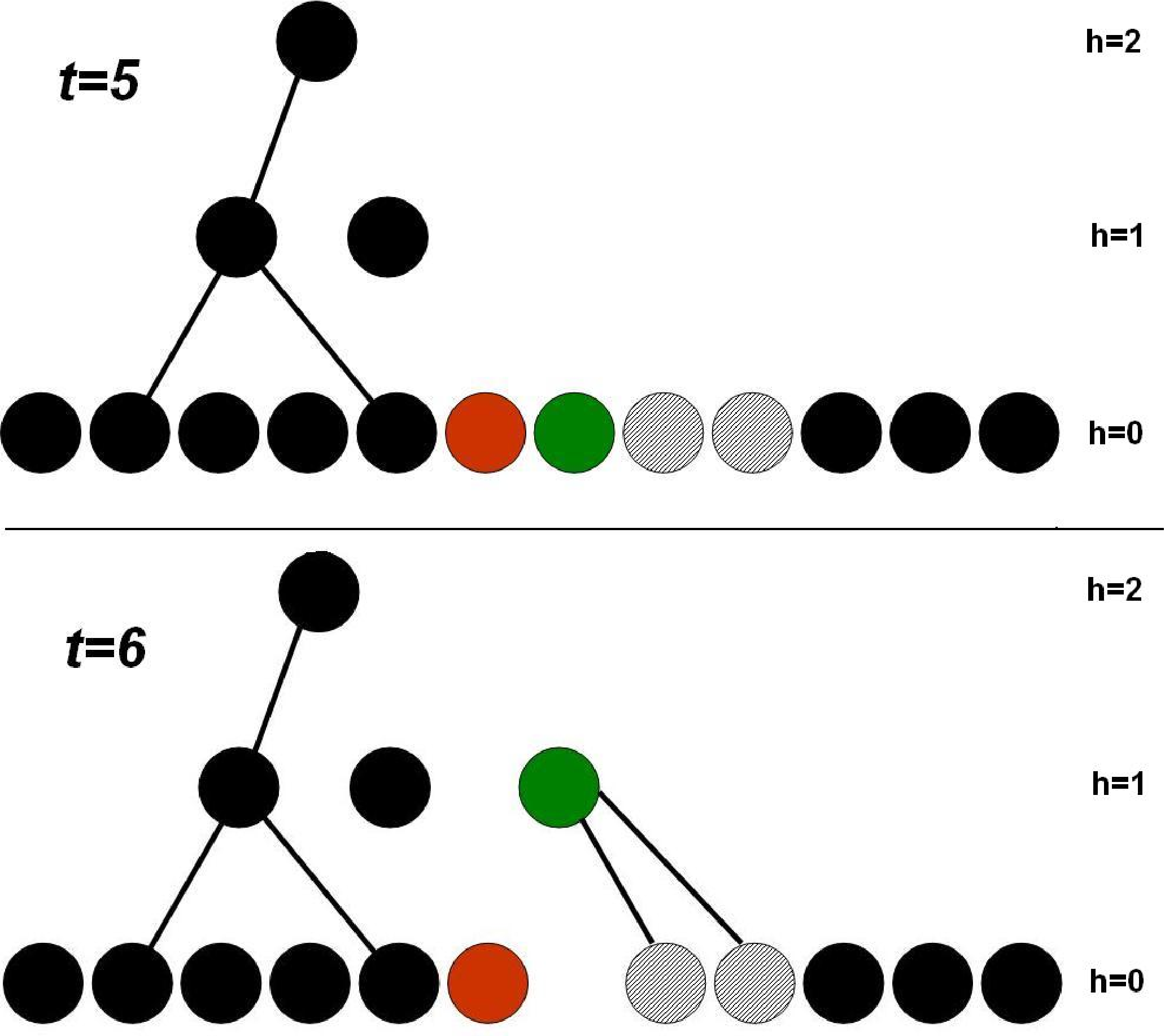,width=0.7\columnwidth}}
\caption{The initial stage of the hierarchy growth. Two consecutive steps of dynamics ($t=5$ and $t=6$). One agent $n=1$ is promoted (green), and gains $\mu=2$ new followers (dashed nodes), and one agent is degraded to $h=0$ (red). In the given example the algorithm selected for degradation an agent from the ground level, in that case agent cannot change their position.}
\label{Fig:step56}
\end{figure} 

\begin{figure}[ht]
\centerline{\psfig{file=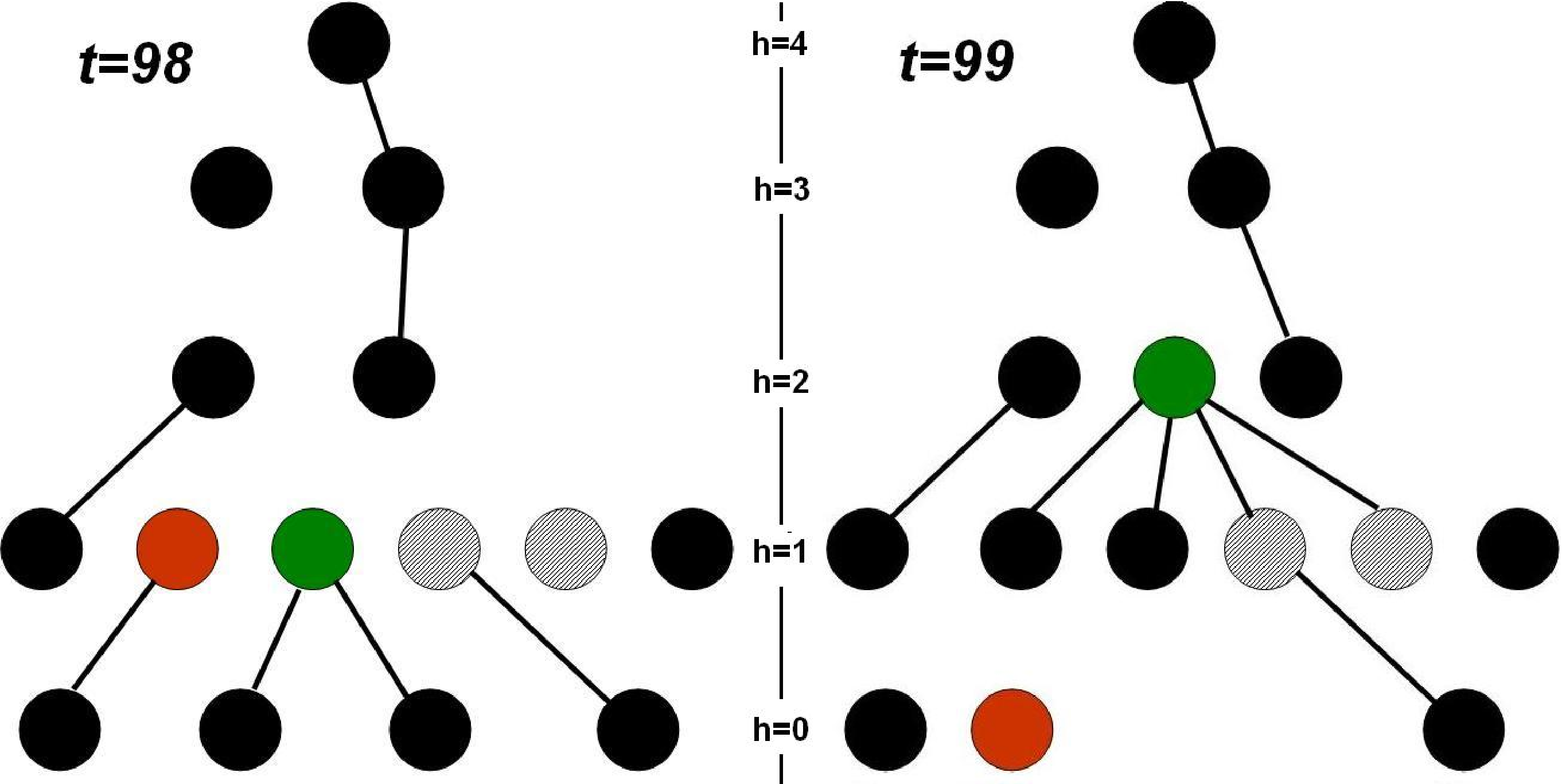,width=0.8\columnwidth}}
\caption{
The stationary state of the hierarchy growth. Two consecutive steps of dynamics ($t=98$ and $99$). One agent $n=1$ is promoted (green), and gains $\mu=2$ new followers (dashed nodes), one agent is degraded to $h=0$ (red). }
\label{Fig:step9899}
\end{figure} 

As previously stated, the proposed model considers a network with a constant number of nodes, while the number of edges and hierarchy levels may vary over time.  Level $h=0$ is referred to as the ground level. Model dynamics is driven by two processes: agents promotions (agents advance to a higher level) and their degradations (agents return to the ground level). It is important to note that these processes are not assumed to be complementary. Degradation does not entail a demotion of an agent (which would merely be a decrease in the hierarchy), but rather the replacement of an agent by a new individual starting from the bottom ($h=0$). It can correspond to the retirement of an old  agent  and the recruitment of a new one, who starts their career from a low-ranking position.

In each time step $n$ agents are promoted and $d$ agents are degraded. In both cases, individual agents are selected randomly from all available agents. When an agent $i$ at a level $h_i$ is promoted, $\mu$ new followers from the same level connect to $i$. These new links represent 'connections' of the promoted agent $i$ to  a set of agents who have chosen $i$ to be their representative. Simultaneously, the promoted agent $i$ and their new followers lose connections with their current leaders.  It is noteworthy that the promoted agent typically had a pre-existing number of followers prior to their promotion, and that all of these followers, in turn, are also promoted, thereby incrementally raising their hierarchical levels by one. This avalanche process describes the character of collective promotions that can be observed in many social, political and business organisations when the rise of the status of an individual positively influences the status of his co-workers (see e.g. \citep{leadership}).   In instances where the level $h_i$ contains fewer than $\mu+1$ nodes, no agent from this level is promoted. Instead, another node (from a different level) is selected for the promotion until this condition is met. 

The degradation of agent $j$ means the loss of all their connections and subsequent transfer to the ground level ($h=0$). It is important to emphasize that, in the context of our model, connections between the nodes at neigbouring hierarchy levels represent certain interactions that result in avalanches of promotions. It is notable that even in the absence of direct promotion, an agent's relevance can be enhanced by the promotion of their leader (or leader's leader, etc.). This promotion avalanche can be visualized as a shift of a connected 'tree' of agents. 

Initially, all agents are situated at the ground level $h=0$. As the system evolves their placements undergo changes, resulting in alterations to their occupation of hierarchy levels and the overall structure of connections. It should be noted that edges can be added, rewired or lost; consequently, the created network will never be fully connected. In this study, we will focus on statistical features of hierarchies in our system and a detailed analysis of topology will be neglected. 

\section{Analytical and numerical results for hierarchy growth}
\label{sec:Results}

\subsection{Topology evolution}
The emergence of new hierarchical levels is the consequence of two separate dynamical processes: namely, the promotion and degradation of nodes. In this study, external forces are not considered. Interactions between agents are represented with links. It is important to note that over time randomly selected nodes undergo changes in their respective hierarchy levels while the overall size of the network remains constant. As previously mentioned, at the initial time step $t=0$ all of the agents occupy level $h=0$ (ground state): $N_0(0)=N$. In each subsequent time step ($t>0$) promotions and degradations of agents are taken into account, resulting in a change in the number of nodes at levels $N_h(t)$. It must be noted that the number of existing levels is also subject to change.

\begin{figure}[ht]
\centerline{\psfig{file=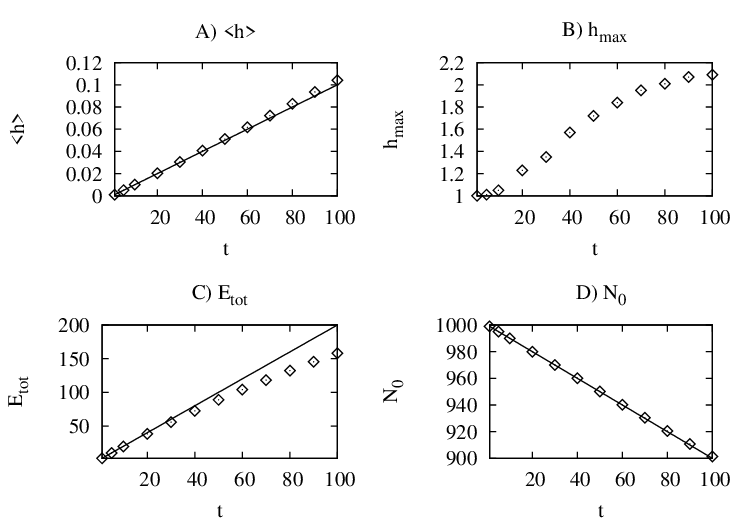,width=0.9\columnwidth}}
\caption{The initial stage of the system evolution for $n=1$, $\mu=2$, $d=1$ and $N=1000$. (a) The average hierarchy level $\left<h\right>(t)$. The line corresponds to the equation $nt/N$. (b) The maximal hierarchy level $h_{max}(t)$. (c) The total number of edges $E_{tot}(t)$. The line corresponds to the equation $n \mu t$. (d) The number of nodes at the ground level $N_0(t)$. The line corresponds to the equation $N-nt$. The results of the computer simulations are averaged over $Q=100$ realizations. }
\label{Fig:m2timepocz}
\end{figure}

\begin{figure}[ht]
\centerline{\psfig{file=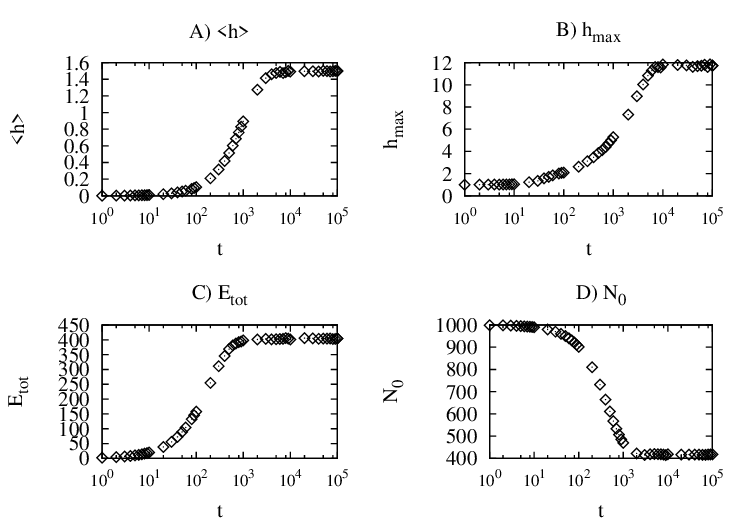,width=0.9\columnwidth}}
\caption{The system's evolution for $n=1$, $\mu=2$, $d=1$ and $N=1000$. (a) The average hierarchy level $\left<h\right>(t)$. (b) The maximal hierarchy level $h_{max}(t)$. (c) The total number of edges $E_{tot}(t)$. (d) The number of nodes at the ground level $N_0(t)$. The results of the computer simulations are averaged over $Q=100$ realizations. }
\label{Fig:m2time}
\end{figure}

The objective of this study is to elicit qualitative and quantitative changes in the hierarchical features of the system that are described by the following macroscopic measures:
\begin{description}
    \item[a)] The average hierarchy level, denoted by $\left<h\right>(t)$;
    \item[b)] The maximal (highest) hierarchy level, denoted by $h_{max}(t)$;
    \item[c)] The total number of links, denoted by $E_{tot}(t)$;
    \item[d)] The number of agents at the ground state $N_0(t)$.
\end{description}  

\begin{figure}[ht]
\centerline{\psfig{file=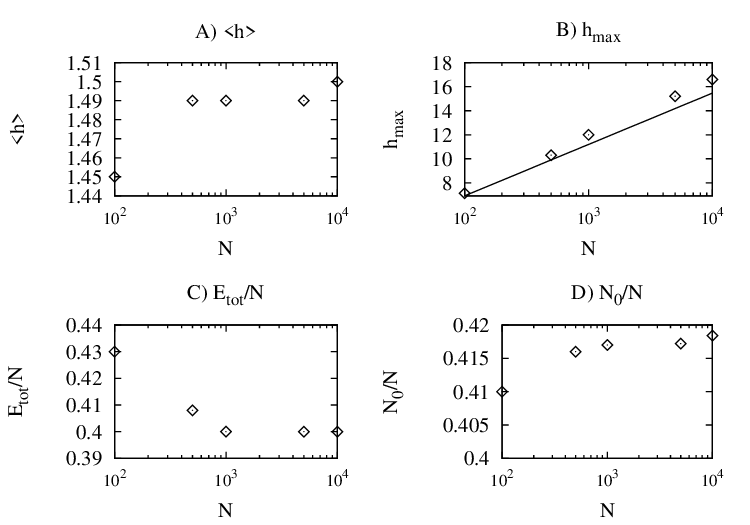,width=0.9\columnwidth}}
\caption{Hierarchical features of the system in the stationary state as a function of $N$ for $n=1$, $\mu=2$, $d=1$. (a) The average hierarchy level $\left<h\right>(N)$. (b) The maximal hierarchy level $h_{max}(N)$. The line corresponds to Eq. (\ref{hmaxeq}). (c) The total number of edges per node $E_{tot}/N(N)$. (d) The fraction of nodes at the ground level $N_0/N(N)$. The results of the computer simulations are averaged over $Q=500$ realizations and for each realization over the last $N$ time steps of the dynamics in the stationary state. The time of one realization is $T=10\cdot N$. }
\label{Fig:m2odN}
\end{figure}

In the early phase of evolution, the system dynamics can be approximated as a process of $n$ promotions from the ground level to $h=1$. Degradations do not affect the system and are essentially negligible (individuals selected for degradations are situated mostly at $h=0$; see schematic depiction in Fig. \ref{Fig:step56}). Consequently, the average hierarchy level is well determined as $nt/N$ (as illustrated in Fig. \ref{Fig:m2timepocz} (a) ). A promoted agent gains $\mu$ followers and the number of links increases by $n\mu$ per step (see Fig. \ref{Fig:m2timepocz} (c) ). The occupation of the lowest level is reduced by $n$ of the promoted nodes in every time step (refer to Fig. \ref{Fig:m2timepocz} (d) ).  This behaviour persists until $t<\frac{N}{n\left(\mu+1\right)}$ (as demonstrated in Figs. \ref{Fig:m2timepocz} and \ref{Fig:m2time}). Subsequent to this phase, the network structure becomes increasingly complex, thereby making analytical description less trivial. 
During this phase the system dynamics is influenced by degradation events and the collective nature of promotions, where the advancement of a single agent can trigger an 'avalanche' of promotions among their followers. The number of followers evolves over time as agents gain new followers and switch their allegiance to different leaders. 
Following a sufficiently long period of time $t\gg N$, the system reaches a stationary state. In other words the network structure (averaged over a large number of realizations) becomes independent of time (see Fig. \ref{Fig:m2time}).  The process of promotions and degradations reaches a state of dynamic equilibrium, resulting in the absence of observable alterations in the hierarchical features of the system. All key metrics -- the average hierarchy level, the maximal hierarchy level, the total number of links, and the number of nodes at the ground level -- remain approximately constant. It is worth noting that, qualitatively the evolution path has a universal character (it does not depend on the system size $N$ if a large enough number of nodes is fed into the system). The above-mentioned quantities for different system sizes in the stationary state are presented in Fig. \ref{Fig:m2odN}. Only the maximal hierarchy level $h_{max}$ grows logarithmically with $N$ (a detailed discussion of this problem is presented in Section \ref{sec:turnieje:hmax}). Our assumption that the existence of a hierarchy level $h$ is equivalent to observing that level at least once means that for a larger number of nodes there is a grater chance of observing higher levels. The occupation of these levels is minute and does not affect the value of the average hierarchy level. 

Based on the system behaviour in time, especially on the fact that for fixed values of the parameters  $n$, $\mu$, $d$ after the transition period there are stationary  values of the macroscopic   system observables, in the following part of the paper we focus on their analytical description.

\subsection{System in the stationary state} \label{sec:bu:analityka}
In this section we focus on finding the distribution of nodes at hierarchy levels  $P_h$ in the stationary state. Since the number of links in the system and the number of nodes at the ground state are independent from time we can find the average number of followers per node in the whole system as 
\begin{equation}
\kappa=\frac{E_{tot}}{N-N_0}.
\label{eqnumkapp}
\end{equation}

\begin{figure}[ht]
\centerline{\psfig{file=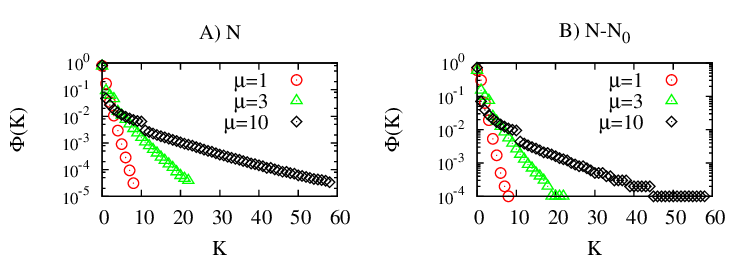,width=\columnwidth}}
\caption{Distributions of promotions avalanches $\Phi(K)$ for (a) all agents in the system and (b) agents at levels $h>0$. Computer simulations are performed for: $n=1$, $\mu=2$, $d=1$, $N=1000$ and are averaged over $Q=500$ realizations and for each realization  over last $N$ time steps of the dynamics in the stationary state. The time of one realization is $T=10^4$.}
\label{Fig:avalanches}
\end{figure}

If this quantity is observed locally, it is equal to $0$ for nodes at $h=0$ (these nodes cannot have followers), and we will assume that for higher levels ($h>0$) it does not depend on $h$. Numerical simulations show that this assumption is true for the majority of levels. 
The number of followers is strictly related to the size of the promotion avalanche $K$, i.e. the total number of indirectly promoted agents (excluding the selected nodes). The value $K$ corresponds to the size of the avalanche \citep{bak} or the mean local reaching centrality \citep{mones}. The distribution of promotion avalanches $\Phi(K)$ is shown in Fig. \ref{Fig:avalanches}.
As can be seen, the probability decreases approximately exponentially with the size of the avalanche $K$. The observed decrease is faster for larger $\mu$. Since in the numerical data $\Phi(0)+\Phi(1)+\Phi(2)>0.9$, in our analytical treatment we neglected the interactions between hierarchy levels that are not in the nearest neighbourhood ($\neq h\pm 1$).

The above approximation allows us to express changes in the number of agents at the ground level $N_0(t)$ by the following processes. During a single promotion event, the occupation of level $h=0$ decreases by $1$ due to the promotion of an agent of this level with the probability $\frac{N_0(t)}{N}$ or by $\kappa$ due to the promotion of an agent of level $h=1$ with the probability $\frac{N_1(t)}{N}$. On the other hand, the degradation of an agent from any level $h>0$ increases $N_0$ by $1$ with the probability $\frac{N-N_0(t)}{N}$. The dynamics can be described by a simple rate equation.
 
\begin{equation}
N_0(t+1)= N_0(t)-n\frac{N_0(t)}{N}+d\frac{N-N_0(t)}{N}-n\kappa(t)\frac{N_1(t)}{N}. \label{N0td}
\end{equation}

Analogously to the case $h=0$, we can write the rate equation for any level $h>0$ by taking into account promotions from levels $h-1$, $h$, $h+1$ and degradations from $h$ as follows.

\begin{eqnarray}
N_h(t+1)=N_h(t)-d\frac{N_h(t)}{N}+n\frac{N_{h-1}(t)}{N}+ \\ \nonumber -n\frac{N_h(t)}{N}+  n\kappa(t)\frac{N_h(t)}{N}-n\kappa(t)\frac{N_{h+1}(t)}{N}. \label{aaaa}
\end{eqnarray}

Let us define $\gamma=\frac{d}{n}$ as the degradation ratio. Then the equations (\ref{N0td}) and (\ref{aaaa}) in a stationary state can be written as 
\begin{equation}
\gamma N-(\gamma+1)N_0 -\kappa N_1=0, 
\label{N0ntdb}
\end{equation}

\begin{equation}
\left(\gamma+1-\kappa\right)N_h-N_{h-1}+\kappa N_{h+1}=0. 
\label{Nhntd}
\end{equation}

Note that Eq. (\ref{Nhntd}) is a recurrence equation with a general solution
\begin{equation}
N_h=N_0\lambda^h,
\label{eqgenNh}
\end{equation}
where $\lambda=\frac{N_{h+1}}{N_h}$ is the ratio of the occupancy of neighbouring levels  and is independent of $h$ (similar to the quantity $\kappa$ this assumption holds for the majority of levels observed in numerical simulations). The total number of nodes in the system is constant and can be written as the sum
\begin{equation}
N=\sum_{h=0}^{h_{max}}N_h=N_0\sum_{h=0}^{h_{max}}\lambda^h,
\label{eqsumNh}
\end{equation}
where $h_{max}$ is the maximal hierarchy level in the system. Assuming $h_{max} \gg 1$ we find
\begin{equation}
\frac{N_0}{N} \approx 1-\lambda.
\label{n0eq}
\end{equation} 

In the stationary state the number of lost and added links is equal, thus $E_{tot}=const$ (see Fig. \ref{Fig:m2time}). If we consider the promotion of $n$ nodes and the degradation of $d$ nodes in a given time step, we can write the change in the total number of links in the system as

\begin{equation}
\Delta E_{tot} =-n\kappa\lambda+n\mu\left(1-\kappa\lambda\right)-d\kappa\lambda-d\kappa
\lambda=0.
 \label{lamkap}
\end{equation}

The successive elements of the above equation correspond to the following processes:
\begin{description}
\item[$(i)$] $-n\kappa\lambda$ $\rightarrow$ promoted nodes lose connections with current leaders,
\item[$(ii)$] $+n{\mu}\left(1-\kappa\lambda\right)$ $\rightarrow$ each promoted agent gains  ${\mu}$ new followers who lose connections to their current leaders,
\item[$(iii)$] $-d\kappa\lambda$ $\rightarrow$ degraded agents lose their connection to their leaders, 
\item[$(iv)$] $-d\kappa\lambda$ $\rightarrow$ degraded agents from level $h>0$ (probability that agent is not at level $h=0$ is $\frac{N-N_0}{N}=\lambda$) lose connections to their followers.
\end{description}

Considering equations (\ref{N0ntdb}) - (\ref{n0eq}), after some algebra one can find analytical forms of the quantities $\kappa$, $\lambda$, $N_0/N$, $E_{tot}/N$.

\begin{equation}
\kappa({\mu},\gamma)=\mu\frac{2\mu+\gamma\mu+2\gamma^2+3\gamma+1}{\left(2\mu+1+2\gamma \right)\left(\mu+1+2\gamma \right)}
 \label{kappafunmd}
\end{equation}

\begin{equation}
\lambda(\mu,\gamma)=\frac{2\mu+1+2\gamma}{2\mu+\gamma\mu+2\gamma^2+3\gamma+1}
\label{lambdafunmd}
\end{equation}

\begin{equation}
\frac{N_0}{N}\left(\mu,\gamma\right)=\gamma\frac{\mu+1+2\gamma}{2\mu+\gamma\mu+2\gamma^2+3\gamma+1}
\label{n0eqmd}
\end{equation} 

\begin{equation}
\frac{E_{tot}}{N}\left(\mu,\gamma\right)=\frac{\mu}{\mu+1+2\gamma}
\label{eqetotmd}
\end{equation}

Next, we insert Eqs. (\ref{n0eqmd}) and (\ref{lambdafunmd}) in Eq. (\ref{eqgenNh}) and define the probability of finding a node at level $h$ as $P_h=\frac{N_h}{N}$. Finally we arrive at
\begin{equation}
P_h=\frac{\gamma\left(\mu+1+2\gamma\right)}{2\mu+\gamma\mu+2\gamma^2+3\gamma+1} \left( \frac{2\mu+1+2\gamma}{2\mu+\gamma\mu+2\gamma^2+3\gamma+1}\right)^h.
\label{eqgenPhd}
\end{equation}

The simple sum can be used to approximate the average hierarchy level
 \begin{equation}
\left<h\right>=\sum_{h=0}^{\infty}\left(h\cdot P_h\right)=\frac{2\mu+2\gamma+1}{\gamma\left(\mu+2\gamma+1\right)}.
\label{eqhavg}
\end{equation}

\begin{figure}[ht]
\centerline{\psfig{file=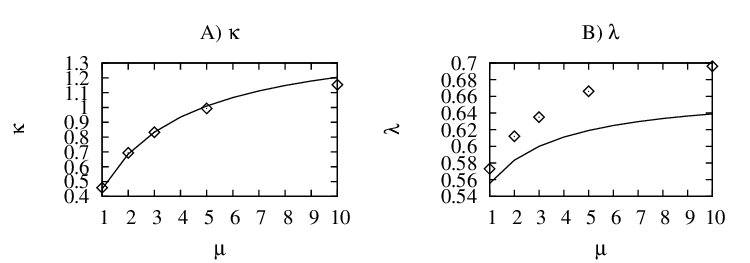,width=\columnwidth}}
\caption{System characteristic quantities in the stationary state as a function of the number of new followers $\mu$, for $n=1$, $d=1$ and $N=1000$. The results of the computer simulations are presented with symbols. (a) The average number of node followers $\kappa(\mu)$. The line corresponds to the equation (\ref{kappafunmd}). (b) The ratio of the number of nodes at two neighbouring levels $\lambda(\mu)$. The line corresponds to the equation (\ref{lambdafunmd}). The results of the computer simulations are averaged over $Q=500$ realizations and for each realization over the last $N$ time steps of the dynamics in the stationary state. The time of one realizations is $T=10^4$.}
\label{Fig:kaplamm}
\end{figure}

 \begin{figure}[ht]
\centerline{\psfig{file=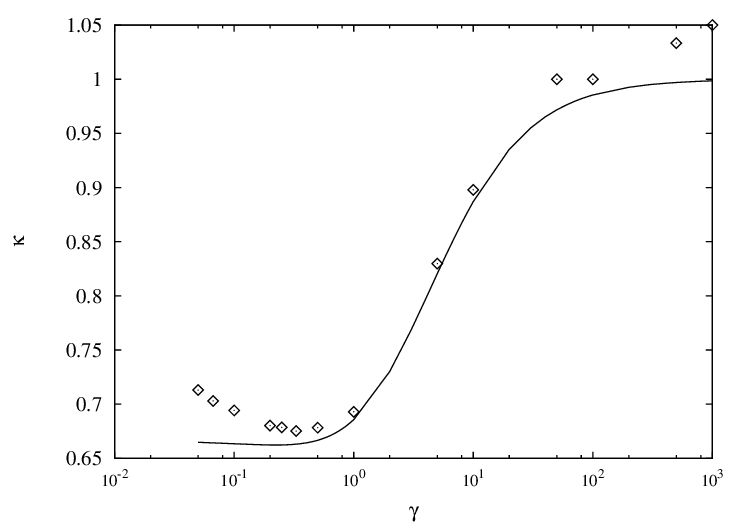,width=0.6\columnwidth}}
\caption{The average number of node followers as a function of degradation ratio $\kappa(\gamma)$ for $\mu=2$ and $N=1000$. The results of the computer simulations are presented with symbols. The line corresponds to Eq. (\ref{kappafunmd}). Points corresponding to $\gamma>1$ are obtained for $n=1$ and $d>1$, $\gamma<1$ corresponds to $d=1$ and $n>1$. The results of the computer simulations are averaged over $Q=500$ realizations and for each realization over the last $N$ time steps of the dynamics in the stationary state. The time of one realizations is $T=10^4$.}
\label{Fig:kappad}
\end{figure}

\begin{figure}[ht]
\centerline{\psfig{file=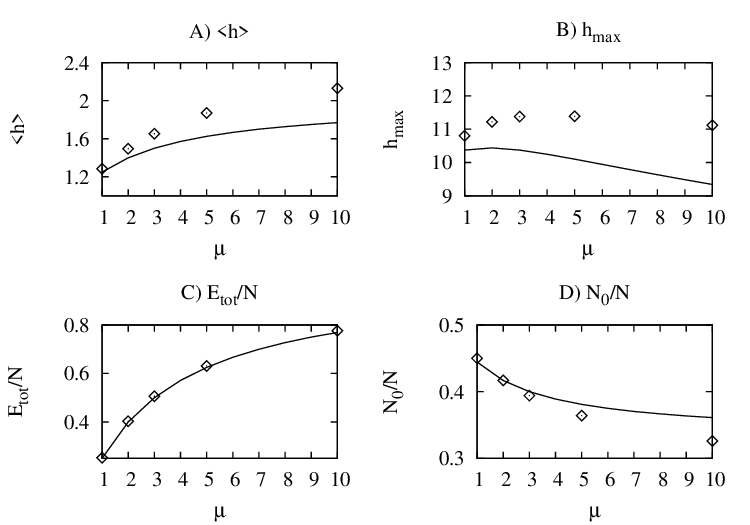,width=\columnwidth}}
\caption{Hierarchical features of the system in the stationary state as a function of the number of new followers $\mu$, for $n=1$, $d=1$ and $N=1000$. The results of the computer simulations are presented with symbols.  (a) The average hierarchy level $\left<h\right>$. The line corresponds to Eq. (\ref{eqhavg}).  (b) The maximal hierarchy level $h_{max}$. The line corresponds to Eq. (\ref{hmaxeq}). (c) The total number of edges per node $E_{tot}/N$. The line corresponds to Eq. (\ref{eqetotmd}). (d) The fraction of nodes at the ground level  $N_0/N$. The line corresponds to Eq. (\ref{n0eqmd}). The results of the computer simulations are averaged over $Q=500$ realizations and for each realization over the last $N$ time steps of the dynamics in the stationary state. The time of one realizations is $T=10^4$.}
\label{Fig:odm1000}
\end{figure}

\begin{figure}[ht]
\centerline{\psfig{file=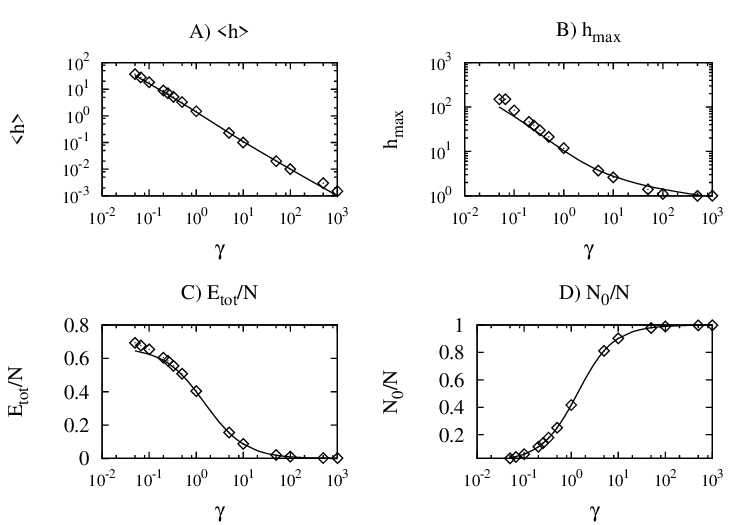,width=\columnwidth}}
\caption{Hierarchical features of the system in the stationary state as a function of the degradation ratio $\gamma$ for $\mu=2$, $N=1000$. The results of the computer simulations are presented with symbols. Points corresponding to $\gamma>1$ are obtained for $n=1$ and $d>1$, $\gamma<1$ corresponds to $d=1$ and $n>1$. (a) The average hierarchy level $\left<h\right>$. The line corresponds to Eq. (\ref{eqhavg}).  (b) The maximal hierarchy level $h_{max}$. The line corresponds to Eq. (\ref{hmaxeq}). (c) The total number of edges per node $E_{tot}/N$. The line corresponds to Eq. (\ref{eqetotmd}). (d) The fraction of nodes at the ground level $N_0/N$. The line corresponds to Eq. (\ref{n0eqmd}). The results of the computer simulations are averaged over $Q=500$ realizations and for each realization over the last $N$ time steps of the dynamics in the stationary state. The time of one realizations is $T=10^4$.}
\label{Fig:m2odd}
\end{figure}

\begin{figure}[ht]
\centerline{\psfig{file=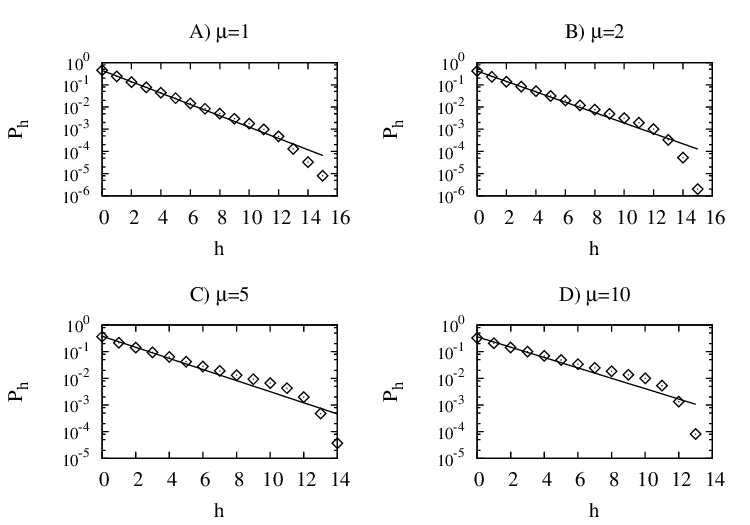,width=\columnwidth}}
\caption{Distributions of nodes at hierarchy levels $P_h$ for different number of new followers $\mu$: (a) $\mu=1$, (b) $\mu=2$, (c) $\mu=5$, (d) $\mu=10$.  The results of the computer simulations for $n=1$, $d=1$ and $N=1000$ are represented by symbols, lines corresponds to Eq. (\ref{eqgenPhd}). The results of the computer simulations are averaged over $Q=500$ realizations and for each realization over the last $N$ time steps of the dynamics in the stationary state. The time of one realizations is $T=10^4$.}
\label{Fig:Ph}
\end{figure}

Comparison of computer simulations and analytical expressions is shown in Figs \ref{Fig:kaplamm} - \ref{Fig:Ph}.

An interesting observation is that the average number of followers $\kappa$ is very low even for a large number of new followers $\mu$ (see Fig. \ref{Fig:kaplamm} (a) ), $\kappa<1$ for $\mu<5$. This effect is caused by the loss of the agent's leader in the course of the promotion process. The quantity $\kappa$ increases almost monotonically with the degradation ratio $\gamma$ (Fig. \ref{Fig:kappad}). The ratio of the number of nodes at neighbouring levels $\lambda$ increases with the number of new followers $\mu$ (with fixed $\gamma$, see Fig. \ref{Fig:kaplamm} (b) ).  When the number of degradations is much smaller than the number of promotions, i.e. $\gamma\rightarrow 0$ the number of nodes at all levels is almost the same, i.e.  $\lambda\rightarrow 1$. A larger number of promotions compared to degradations makes the system more heterogeneous. Intuitively, the fraction  of nodes at the ground level $N_0/N$ should decrease with $\mu$  and increase with $\gamma$ and it has been confirmed by numerical simulations and analytical solutions (see Figs. \ref{Fig:odm1000} (d) and \ref{Fig:m2odd} (d) ). The opposite behaviour is observed with respect to the number of edges per node (see Figs. \ref{Fig:odm1000} (c) and \ref{Fig:m2odd} (c) ). A large number of degradations results in a smaller number of links, leading to a higher occupancy at the ground level. The average hierarchy level increases with $\mu$ (see Fig. \ref{Fig:odm1000} (a) ) and decreases with $\gamma$ (see Fig. \ref{Fig:m2odd} (a) ).   The distributions $P_h$ observed in numerical simulations (Fig. \ref{Fig:Ph}) show the exponential decay of hierarchy level occupancy predicted by Eq. (\ref{eqgenPhd}). We observe a more rapid decrease for higher $\mu$, where the differences between the occupation of levels are larger.

\subsection{Maximal hierarchy level} \label{sec:turnieje:hmax}
 During the computer simulations, the maximal hierarchy level was defined as the highest $h$ that was observed at least once throughout all steps and all realizations of the dynamics. This means that the said quantity can be considered as a kind of fluctuation and its size tends to grow along $N$. In the stationary state we do not observe the emergence of new levels. According to the dynamic rules, a new level cannot emerge if the occupation of the existing maximal level does not exceed $\mu$ nodes. In order to find the analytical form of $h_{max}$, we assumed that the average number of nodes at the highest level should be $\frac{\mu+1}{2}$. This corresponds to a  situation where there can be  $N_{h_{max}}=1,2,\ldots,\mu$ agents at the maximum level, and all these  cases are possible with equal probability. Substituting $P_{h_{max}}=\frac{\mu+1}{2N}$ into Eq. (\ref{eqgenNh}) we get 

\begin{equation}
h_{max}=\frac{\ln{\frac{\mu+1}{2 N}}-\ln\left(1-\lambda\right)}{\ln\left(\lambda\right)}.
\label{hmaxeq}
\end{equation}

The logarithmic growth of $h_{max}$ with the system size $N$ given by Eq. (\ref{hmaxeq})  is confirmed by the numerical simulations presented in in Figs. \ref{Fig:m2odN} (b), \ref{Fig:odm1000} (b) and \ref{Fig:m2odd} (b).

\subsection{Critical degradation ratio} \label{sec:bu:ogolniej}
Up to now we have shown that for fixed system parameters ($\gamma=d/n$, $N$ and $\mu$) the system reaches the stationary state and the limited value of $h_{max}$. One question, of course, is whether there is a critical value of the degradation ratio   $\gamma^*$, for which the growth of hierarchies is not observed, i.e. $h_{max}=1$. This condition corresponds to a situation where on average only one agent can be found at a level $h>0$ (and the remaining $N-1$ agents occupy the ground level, $h=0$).

\begin{equation}
N_0=N-1.
\label{warunekdgw}
\end{equation}

Considering the above conditions and Eqs. (\ref{lambdafunmd}), (\ref{n0eqmd}), (\ref{eqgenPhd}) we get

\begin{eqnarray}
\gamma^*(\mu,N)=\frac{\sqrt{\mu^2-10\mu+12\mu N+\left(2N-1\right)^2}}{4} + \frac{2N-\mu-3}{4}\xrightarrow{\mu\ll N, N\rightarrow\infty} N.
\label{dgwmN}
\end{eqnarray}

Equation (\ref{dgwmN}) means that hierarchical structures are created even when the number of degradations $d$ is much higher as compared to the number of promotions $n$. Moreover, for large systems the critical degradation ratio is comparable to the number of nodes, $N$.  

\section{Leaders without followers - individual promotions} \label{sec:bu:mu0}

So far, only collective promotions have been considered, i.e. each promoted agent has caused the promotion of their followers and the subsequent followers. A question to be addressed below is how much this affects the system behaviour. To answer this, we performed numerical simulations and found an analytical solution for a scenario where promoted agents have no followers $\mu=0$ (i.e. individual promotions) and thus $\kappa=0$ and $E_{tot}=0$. The rate equations (\ref{N0td}) and (\ref{aaaa}) lose elements with $\kappa$ and we get

\begin{equation}
\frac{N_0(t)}{N}= \frac{1}{n+d}\exp(-\frac{n+d}{N}t)+\frac{d}{n+d} \label{N0tdm0}
\end{equation}

\begin{figure}[ht]
\centerline{\psfig{file=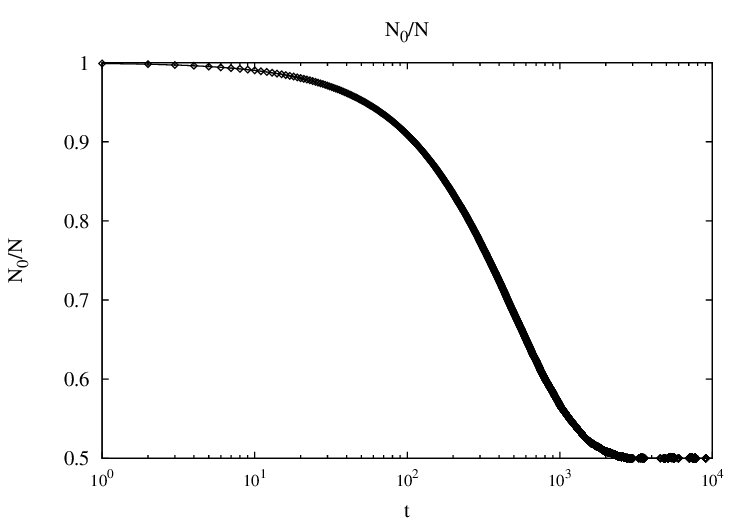,width=0.6\columnwidth}}
\caption{The fraction of nodes at the ground level in time  $N_0(t)/N$. The results of the computer simulations for $n=1$, $\mu=0$, $d=1$ and $N=1000$ are represented with symbols and are averaged over $Q=500$ realizations. The line corresponds to Eq. (\ref{N0tdm0}). }
\label{Fig:N0}
\end{figure}

For $t\rightarrow\infty$ the first term in Eq. (\ref{N0tdm0})  goes to zero and the stationary solution is only provided by the second term. Similarly it is possible to obtain the formula for $N_h(t)$. The number of nodes at the ground level is higher for individual promotions than for collective promotions (see Figs \ref{Fig:m2time} (d) and \ref{Fig:N0}). This effect confirms that collective promotions produce steeper hierarchical systems.

Similarly to the $\mu>0$ case, we can write the stationary form of the rate equations describing $N_0$ and $N_h$ and find the ratio between the number of nodes at neighbouring levels as

\begin{equation}
\lambda=\left(\gamma+1\right)^{-1}. 
\label{lamfungam}
\end{equation}

Now using Eq. (\ref{lamfungam}) we can obtain the number of nodes at  level $h$ as

\begin{equation}
\frac{N_h}{N}=\frac{\gamma}{\gamma+1}\left(\gamma+1\right)^{-h}.
\label{N0m0g}
\end{equation}

The average hierarchy level for $\mu=0$ (using eqs (\ref{eqhavg}) and (\ref{N0m0g})) is given by

\begin{equation}
\left<h\right>\left(\gamma\right)=\frac{1}{\gamma},
\label{eqhavgm0}
\end{equation}

which is simply the ratio of the number of promoted agents to the number of degraded agents. It is worth to note that Eq. (\ref{lamfungam}) can be also directly obtained as a limiting case of Eq. (\ref{lambdafunmd}) by setting $\mu=0$.

According to Eq. (\ref{N0m0g}) when $\gamma\gg1$ (the number of degradations is much higher than the number of promotions) the occupancy of level $h=0$ tends to $N$. In the opposite case, when $\gamma\ll 1$ (the number of degradations is negligible compared to the number of promotions)  $N_0\rightarrow 0$. These results are in line with intuition. In close similarity to the previous Section we can find a critical degradation ratio  $\gamma^*$ above which hierarchies do not emerge
\begin{equation}
\gamma^*=N-1.
\end{equation}

\begin{figure}[ht]
\centerline{\psfig{file=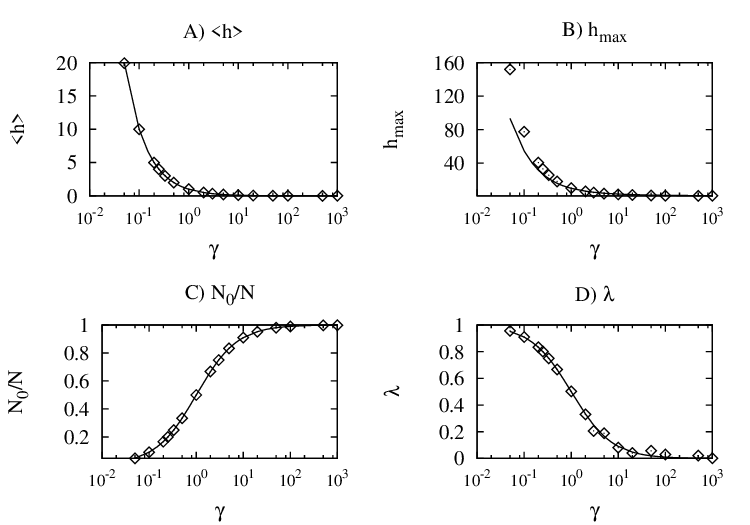,width=\columnwidth}}
\caption{Hierarchical features of the system in the stationary state as a function of the degradation ratio $\gamma$ for $\mu=0$, $N=1000$. The results of the computer simulations are represented with symbols. Points corresponding to $\gamma>1$ are obtained for $n=1$ and $d>1$, $\gamma<1$ correspond to $d=1$ and $n>1$. (a) The average hierarchy level $\left<h\right>(\gamma)$. The line corresponds to Eq. (\ref{eqhavgm0}).  (b) The maximal hierarchy level $h_{max}(\gamma)$. The line corresponds to Eq. (\ref{hmaxeq}). (c) The fraction of nodes at the ground state $N_0/N$. The line corresponds to Eq. (\ref{N0m0g}). (d) The ratio of number of nodes at two neighbouring levels $\lambda=\frac{N_{h+1}}{N_h}$. The line corresponds to Eq. (\ref{lamfungam}).  The results of the computer simulations are averaged over $Q=500$ realizations and for each realization over the last $N$ time steps of the dynamics in the stationary state. The time of one realization is $T=10^4$.}
\label{Fig:m0odd}
\end{figure}

The results of the computer simulations and the theoretical treatment are shown in Fig.\ref{Fig:m0odd}. The qualitative behaviour of our system for  $\mu=0$ is similar to the previous results for the case $\mu>0$ (see Fig. \ref{Fig:m2odd}).  Quantitative differences can be noted for both scenarios. When $\gamma=0.05$, the average hierarchy level is $\left<h(\mu=0,\gamma=0.05)\right>\approx 20$, while $\left<h(\mu=2,\gamma=0.05)\right>\approx 37$. As can be seen, the decrease in the average hierarchy level is almost doubled if we take into account the absence of followers in the system.  Similar differences can be observed for the fraction of nodes at the ground level. In the case of $\mu=0$ and $\gamma=0.05$ at level $h=0$ there are $N_0\approx 0.05N$ agents, while for $\mu=2$ there are almost twice as few ($\approx 0.028$). This shows that the process of hierarchy growth is more pronounced when we consider interactions between individuals (links between nodes). 

\section{Discussion and Conclusions}
We have demonstrated that the process of hierarchy growth in a simple  bottom-up approach converges to a stationary state characterized by  an exponential decay of agent distribution  with the hierarchy  level $h$, as $N_h \sim  \lambda^h$ (where the scaling ratio $\lambda <1$). This equilibrium persists despite continuous system perturbations through collective promotions and individual degradations. Our finding is supported by extensive numerical simulations and validated through analytical results derived from rate equations. Notably, our results are consistent with observed patterns of hierarchical organization in several real-worlds systems, including hunter-gatherers \citep{huntergatherer}, mammals \citep{mammals}, and players of the online game \textit{Pardus} \citep{fuchs}. Our model, like these empirical studies, shows that bottom-up self-organization naturally leads to hierarchies with exponential scaling in group sizes across levels. This convergence underscores that hierarchical organization is an intrinsic and essential process for structuring complex interactions in both human and animal societies.

Interestingly, the distribution of the occupancy of hierarchy levels is reminiscent of the Boltzmann distribution of energy level occupancy where the hierarchy levels $h$ correspond to the energy levels and $-\log(\lambda)$ to the inverse temperature. The scaling ratio  $\lambda$  of agent distribution $N_h$  increases with the number of agent's initial followers ($\mu$), i.e. a higher system temperature corresponds to a larger number of new followers of every promoted agent. 

The stationary state emerges when the system age greatly exceeds its size ($t\gg N$). In this regime, key metrics -- including the average hierarchy level $\left<h\right>$, the total number of edges per agent $E_{tot}/N$, and the fraction of nodes at the ground level $N_0/N$ -- become independent of the system size (beyond a certain threshold). These observables depend on the system parameters, i.e. the number of new followers $\mu$ and the degradation ratio $\gamma=\frac{d}{n}$, where $d$ and $n$ represent the number of degradations and the number of promotions per time step, respectively. While the  system size does not affect these metrics, it does influence the maximal hierarchy level $h_{max}$ which grows logarithmically with $N$  although the occupation of  $h_{max}$ is small and does not impact $\left<h\right>$.

The stationary values of $\left<h\right>$,  $h_{max}$ and  $E_{tot}/N$,  increase  with $\mu$ and decrease with $\gamma$. The fraction of nodes at the ground level exhibits the opposite behaviour, namely decreasing with $\mu$ and increasing with $\gamma$. To better understand the collective nature of the promotions, we calculate the branching factor $\kappa$, defined as the average number of followers of the agent. As expected, it increases with  $\mu$, but it is worth noting that a significant number of initial followers ($\mu>5$) is required to reach a branching factor $\kappa>1$. Interestingly, it shows a non-monotonic relationship with $\gamma$. For $\gamma<1$ there is almost no change in $\kappa$, while for $1<\gamma<100$ we observe rapid increase in the branching factor, but it saturates around the value of $1$ for $\gamma\approx 100$.  

Hierarchical structures persist even when there are no followers ($\mu=0$). In this case, the scaling ratio becomes $\lambda =\left(1+\gamma\right)^{-1}$ and the average hierarchy level becomes $\left<h\right>=1/\gamma $. Since this value is lower than for $\mu>0$  it leads to the conclusion that collective promotions reinforce hierarchy growth.

In our previous work \citep{pre} we examined models of hierarchical networks that grow from the top down, where newcomers seek positions close to the highest level based on limited knowledge of the network structure. We found that the availability of information could constrain the growth of the hierarchy. The bottom-up emergence scenario considered here appears to be more robust, with new levels only ceasing to appear when the ratio of degradations to promotions approaches the size of the system.

\section*{Acknowledgements} 
The research leading to these results has received funding from the European Union Seventh Framework Programme (FP7/2007-2013) under grant agreement no 317534 (the Sophocles project) and from the Polish Ministry of Science and Higher Education grant 2746/7.PR/2013/2. 
The work was also partially supported as  RENOIR Project by the European Union Horizon 2020 research and innovation programme under the Marie Sk\l odowska-Curie grant agreement No 691152  and by Ministry of Science and Higher Education (Poland), grant Nos. 34/H2020/2016,   329025/PnH /2016. and National Science Centre, Poland Grant No.  2015/19/B/ST6/02612.
A.C. was also supported by Warsaw University of Technology within the Excellence Initiative: Research University (IDUB) programme.
J.A.H. was partially supported by  a Grant from The Netherlands Institute for Advanced Study in the Humanities and Social Sciences (NIAS)the Russian Scientific Foundation, Agreement \#17-71-30029 with co-financing of Bank Saint Petersburg and by POB Research Centre Cybersecurity and Data Science of Warsaw University of Technology within the Excellence Initiative Program—Research University (IDUB). 

\appendix

\bibliographystyle{elsarticle-num-names} 
\bibliography{bibliography}





\end{document}